\documentstyle[preprint,eqsecnum,aps]{revtex}

\newcommand{\extraspace}{\addtolength{\abovedisplayskip}{2mm}
                        \addtolength{\belowdisplayskip}{2mm}
                        \addtolength{\abovedisplayshortskip}{2mm}
                        \addtolength{\belowdisplayshortskip}{2mm}}
\newcommand{\be}{\begin{equation}\extraspace}
\newcommand{\ee}{\end{equation}}
\newcommand{\bea}{\begin{eqnarray}\extraspace}
\newcommand{\eea}{\end{eqnarray}}

\newcommand{\nonu}{\nonumber \\[2mm]}

\newcommand{\half}{\scriptsize{\frac{1}{2}}}
\newcommand{\tr}{{\rm tr}}

\begin{document}

\title{\hfill {\rm Oct. 1995} \\[15mm] \large
Thermodynamic Bethe Ansatz for $N\!=\!1$ Supersymmetric Theories}
\author{ M. Moriconi}
\address{ Physics Department, Princeton University
\\Jadwin Hall, Princeton, NJ 08544, U.S.A.}

\author{K. Schoutens}
\address{ Institute for Theoretical Physics, University of Amsterdam\\
Valckenierstraat 65, 1018 XE Amsterdam, The Netherlands}

\maketitle
\begin{abstract}
We study a series of $N\!=\!1$ supersymmetric integrable particle theories
in $d=1+1$ dimensions. These theories are represented as integrable
perturbations of specific $N\!=\!1$ superconformal field theories.
Starting from the conjectured $S$-matrices for these theories, we develop
the Thermodynamic Bethe Ansatz (TBA), where we use that the 2-particle
$S$-matrices satisfy a free fermion condition. Our analysis proves a
conjecture by E.~Melzer, who proposed that these $N\!=\!1$
supersymmetric TBA systems are ``folded'' versions of $N\!=\!2$
supersymmetric TBA systems that were first studied by P.~Fendley and
K.~Intriligator.
\end{abstract}

\vfill

\noindent PUPT-1572

\noindent ITFA-95-17

\noindent hep-th/9511008

\newpage
\section{Introduction}
Integrable quantum field theories in $d=1+1$ dimensions are more than
just ``theoretical laboratories". Apart from being very rich mathematical
structures, they have proven to be of interest for a variety of
problems in theoretical physics, ranging from string theory to
statistical mechanics lattice models and problems in condensed matter
theory. At the technical level, the magic of integrable field theories
can be traced back to the existence of an infinite number of non-trivial
conserved charges \cite{zz}. These imply the conservation of individual momenta
in many-particle collisions and the factorizability of the $S$-matrix.
The two-body $S$-matrix satisfies the Yang-Baxter equation, in addition
to the usual properties of unitarity, crossing symmetry and analyticity.
In many cases, these properties of the $S$-matrix are so restrictive
that one may set up a bootstrap program and determine the exact $S$-matrix,
up to the famous ``CDD ambiguity" \cite{cdd}.

One special class of integrable field theories, that we will be studying
in this paper, are the integrable massive perturbations of conformal
field theories.
Such theories are obtained by starting with a conformal field
theory (CFT) and perturbing it by a relevant operator that preserves
integrability. We will see the technicalities later.
The ultraviolet (UV) properties of the CFT will not be affected by the
perturbation, but the infrared (IR) behavior will change. The modified
IR behavior can be conformal or massive (finite correlation
length). In the latter case, the IR behavior is given by a massive
particle theory with factorizable scattering and one may employ the
above-mentioned bootstrap program to identify the exact $S$-matrix.

Once the $S$-matrix of an integrable field theory is known, one
may study its thermodynamics by following a procedure called
Thermodynamic Bethe Ansatz (TBA) \cite{zama2,klamel1}.
In the context of applications
in condensed matter problems (such as the scattering of edge
currents in the fractional quantum Hall effect, \cite{fls}), the
TBA results offer concrete predictions that can be tested in
experiments or simulations. Another nice feature is that, among
the quantities that can be computed by the TBA are the central charge
and the scaling dimensions of the UV limit of the theory. In this way,
the TBA may be used as a check on the validity of a conjectured
$S$-matrix for a perturbed CFT. The name of the TBA technique,
which will be reviewed briefly in section III, is a little misleading,
since the TBA is an exact non-perturbative procedure and not a
trial-and-error method.

The theories that we study in this paper are examples of
integrable field theories with supersymmetry. Adding supersymmetry
is of course natural in the context of string theory applications,
but even in the context of statistical mechanics lattice models
supersymmetry is meaningful. For example, the tricritical Ising model
in two dimensions \cite{fsz} is an example of a physical system that realizes
superconformal symmetry. If we start from a superconformal
field theory and choose a perturbation that preserves supersymmetry
we end up with a massive theory with supersymmetry. In such a
case supersymmetry may be added to the bootstrap ingredients,
and one may try to identify exact $S$-matrices by exploiting both
supersymmetry and factorizability.

In this paper we will study the TBA for a series \cite{schoutens1} of
$N\!=\!1$ supersymmetric scattering theories with $2n$ particles,
$n$ some positive integer, arranged in $n$ supermultiplets of one
bosonic and one fermionic particle each, with masses
labeled by an integer $k$ and given by
\be
m_{k}={\sin(k \beta \pi) \over \sin(\beta \pi)} \ , \ \label{intro1}
\ee
where $\beta=1/(2n+1)$ and $k=1,2, \cdots ,n$. It has been conjectured
\cite{schoutens1} that these scattering theories correspond to an
integrable perturbation of specific superconformal field theories of
central charge $c_n=-3n(4n+3)/(2n+2)$. Note that this is a series of
non-unitary theories.

One important observation is that the $S$-matrix for these particular
theories is
non-diagonal, since we can have, for example, a scattering process of the
form $b_{i}b_{j} \rightarrow f_{i}f_{j}$, where $b_{i} (f_{i})$ is a
boson (fermion) in the $i$-th supermultiplet. This makes the TBA analysis
more difficult than in the case of diagonal scattering, where the
thermodynamic limit can be studied via a set of coupled integro-differential
equations. In the non-diagonal case we will have to diagonalize a tranfer
matrix in order to obtain a tractable system of equations.

In general, the transfer matrix that features in the TBA for non-diagonal
scattering theories is identical to the transfer matrix of a
corresponding statistical mechanics lattice model, which is obtained
by interpreting the two-body $S$-matrix elements as Boltzmann weights.
In the case of the above $N\!=\!1$ supersymmetric theories, this lattice
model is of
eight-vertex type, and one might expect that the analysis would be
quite involved. However, it has been observed that in a rather general
setting \cite{shanko} the eight-vertex models that correspond to
$N\!=\!1$ scattering
matrices satistfy a technical condition callled the free fermion condition.
This case is no exception, and we shall use this special property to
determine the eigenvalues of the transfer matrix and to complete
the TBA analysis.

One of the motivations for us for deriving our $N\!=\!1$ supersymmetric
TBA systems has been a proposal made by E.~Melzer in \cite{melzer1}.
He conjectured that the $N\!=\!1$ TBA systems can be derived by starting
from scattering theories with $N\!=\!2$ supersymmetry, and twice as many
particles, and applying a procedure called ``folding'' on these
$N\!=\!2$ TBA system. These $N\!=\!2$ TBA systems were first studied by
P.~Fendley and K.~Intriligator in \cite{fenit1}. Melzer checked that the
``folded" TBA systems had all the right properties to be related to the
$N\!=\!1$ superconformal theories that underly the $N\!=\!1$ $S$-matrices,
but was unable to give a direct derivation starting from the $N\!=\!1$
$S$-matrices. In this paper we fill in this missing link and confirm
Melzer's conjecture by showing that the true $N\!=\!1$ TBA systems
are indeed identical to the folded $N\!=\!2$ systems of \cite{melzer1}.

In earlier work, the bosonic scattering theories that underly the
$N\!=\!1$ and $N\!=\!2$ supersymmetric scattering theories that we just
discussed have been related by a similar folding,
both at the level of the TBA and
at the level of the $S$-matrices \cite{zama2,klamel1}.
Our results in this paper
clearly suggest that it should be possible to find a ``folding
relation'' between the $N\!=\!1$ and $N\!=\!2$ supersymmetric theories,
directly at the level of
the $S$-matrices. We will address this issue in a forthcoming
publication \cite{msII}.

This paper is organized as follows. In section II we review the formalism
for supersymmetric particle theories with factorizable scattering
and we introduce a specific series of $N\!=\!1$ supersymmetric theories.
In section III we discuss the TBA technique and highlight some aspects
that will be relevant later. In sections IV and V we obtain
the TBA system for the $N\!=\!1$ supersymmetric theories and verify the
UV and IR limits.
In section VI we present Melzer's folded $N\!=\!2$ systems and demonstrate
that they agree with our $N\!=\!1$ systems. Our conclusions are formulated
in section VII.

% section II
\section{Supersymmetric Particle Theories in 1+1 dimensions with
         Factorizable Scattering}

In this section we briefly introduce our $n$-component
supersymmetric scattering theories and establish some notation.
For a more detailed discussion, see \cite{schoutens1}.

We are interested in perturbed superconformal field theories
decribed by an action of the form
\be
S_{\lambda}= S + {\lambda} \int ({\overline G_{-{1 \over 2}}
G_{-{1 \over 2}} \phi_{h,h}}) \, d^{2}x \ , \ \label{sufa1}
\ee
where $S$ is the action of the unperturbed theory,
which we take to be the minimal superconformal model ${\cal M}(2,4n+4)$ of
central charge $c_n = -3n(4n+3)/(2n+2)$. The perturbation contains
$\phi_{h,h}$, which is a primary field in the Neveu-Schwarz sector
of the left and right-chiral superconformal algebras.
The perturbations in (\ref{sufa1}) are manifestly supersymmetric.
If we consider the case with $h=h_{(1,3)}$ we obtain an integrable theory.
This is what we meant by massive deformations of CFT's in the introduction.
It can be shown by means of a slight variation of Zamolodchikov's counting
argument \cite{zama1} that for this perturbation there is at least one
non-trivial integral of motion, which ensures the factorizability of
the $S$-matrix. Since the perturbation has the dimension of action, we are
automatically introducing a scale in our model and the resulting theory is
not conformal invariant. On the other hand, in the ultra-violet (UV) limit
we obtain a (super)conformal theory. This UV limit is simply the
theory described by $S$ (the unperturbed theory) since in the
limit of extremely high energy scattering, the particles do not ``see" any
finite energy scale. The infra-red (IR) regime of the theory is completely
described by a factorizable scattering theory, which, according to the
conjecture in \cite{schoutens1} can be described as follows.

The IR scattering theory contains $2n$ massive particles,
arranged in $n$ supermultiplets $(b_i,f_i)$, with masses $m_i$ as
given in (\ref{intro1}).
We denote by $A_{i}(\theta_i)$ be any particle, boson or fermion,
from any multiplet, with asymptotic momentum $p^0_i=m_i\cosh(\theta_i)$
and $p^1_i=m_i\sinh(\theta_i)$. Multi-particle states are then written as
\be
|A_{i_1}(\theta_{i_1})A_{i_2}(\theta_{i_2})...A_{i_N}(\theta_{i_N})\rangle_
{in(out)}\ , \
\label{sufa2}
\ee
where $\theta_{i_1} \geq \theta_{i_2} \geq \ldots \geq \theta_{i_N}$
for in-states and the other way around for out-states.

The two-body $S$-matrix will carry two labels, $i$ and $j$, that tell us
which supermultiplets the particles in the scattering belong to. From
Lorentz invariance it is easy to see that the two-body $S$-matrix can
depend only on the rapidity difference $\theta=\theta_i-\theta_j$.
The complete two-body $S$-matrix is written as a product of two pieces
\be
S^{[ij]}(\theta)=S^{[ij]}_{BF}(\theta) \ S^{[ij]}_{B}(\theta)\ \ , \
\label{sufa3}
\ee
where $S^{[ij]}_{BF}$ is a piece of the $S$-matrix that mixes bosons and
fermions and $S^{[ij]}_{B}$ is a bosonic $S$-matrix that acts only in the
bosonic sector.

For the model at hand, the bosonic piece $S^{[ij]}_B$ is the
$S$-matrix found by P.~Freund, T.~Klassen and E.~Melzer in
\cite{fkm} and given by
\be
S^{[ij]}_B(\theta)=-F_{|i-j|\beta}(\theta) \left[
  F_{(|i-j|+2)\beta}(\theta) \ldots
  F_{(i+j-2)\beta}(\theta) \right]^2 F_{(i+j)\beta}(\theta)\ , \
\label{sufa4}
\ee
with $F_{\alpha}(\theta)={{\sinh(\theta)+i\sin(\alpha \pi)} \over
{\sinh(\theta)-i\sin(\alpha \pi)}}$.
In a basis given by $|b_{i}b_{j}\rangle$, $|b_{i}f_{j}\rangle$,
$|f_{i}b_{j}\rangle$, $|f_{i}f_{j}\rangle$ the $S^{[ij]}_{BF}$
piece is of the form
\be
S_{BF}^{[ij]}(\theta)= f^{[ij]}(\theta)
\left(\begin{array}{cccc}
      1-t \widetilde{t} & 0 & 0 & -i(t+\widetilde{t})\\
      0 & -t+\widetilde{t} & 1+t\widetilde{t} & 0\\
      0 & 1+t\widetilde{t} & t-\widetilde{t} & 0\\
      -i(t+\widetilde{t}) & 0 & 0 & 1-t \widetilde{t}
      \end{array}\right)\ + g^{[ij]}(\theta)
\left(\begin{array}{cccc}
      1 & 0 & 0 & 0 \\
      0 & 1 & 0 & 0 \\
      0 & 0 & 1 & 0 \\
      0 & 0 & 0 & -1
      \end{array}\right)\ \ , \
\label{sufa5}
\ee
where $t=\tanh((\theta+\log(m_{i}/m_{j}))/4)$ and
$\widetilde{t}=\tanh((\theta-\log(m_{i}/m_{j}))/4)$.
The functions $f^{[ij]}(\theta)$ and $g^{[ij]}(\theta)$ are related by
\be
f^{[ij]}(\theta)={\alpha \over 4i} {\sqrt{m_{i}m_{j}}}
      \left [
        {2 \cosh(\theta/2)+(\rho^{2} + \rho^{-2})}
        \over {\cosh(\theta/2) \sinh(\theta/2)}
      \right ] g^{[ij]}(\theta)\ , \ \label{sufa6}
\ee
where $\rho=(m_i/m_j)^{1/4}$ and $\alpha=-\sin(\beta \pi)$.
This specific form of the Bose-Fermi $S$-matrix $S^{[ij]}_{BF}$ is almost
completely fixed by $N\!=\!1$ supersymmetry
(which allows only two free functions
$f^{[ij]}(\theta)$ and $g^{[ij]}(\theta)$) and by the Yang-Baxter equation,
which fixes the ratio of $f^{[ij]}(\theta)$ and $g^{[ij]}(\theta)$ up to
one free constant $\alpha$. The value of $\alpha$ is then dictated by
consistency with the bound-state structure induced by the bosonic factor
$S^{[ij]}_B$ in (\ref{sufa4}).

The function $g^{[ij]}(\theta)$ is constrained by the conditions of
analyticity, crossing symmetry and unitarity,
\bea
{\rm analyticity:}& g^{[ij]}(-\theta)=g^{[ij]*}(\theta),
\nonu
{\rm crossing \, symmetry:}& g^{[ij]}(i \pi -\theta)=g^{[ij]}(\theta),
\label{sufa7}
\\[2mm]
{\rm unitarity:}& g^{[ij]}(\theta)g^{[ij]}(-\theta)=
\nonu
   & \left [1+ \alpha^2 m_{i}m_{j}
           \left ( {\sinh^2(\theta/2)+{1 \over 4} ( \rho^{2} + \rho^{-2})^{2}}
           \over {\cosh^{2}(\theta/2) \sinh^{2}(\theta/2)}
           \right)
          \right]^{-1} . \nonumber
\eea
An explicit integral expression for $g^{[ij]}(\theta)$, which will be
important for the TBA analysis, can be found as follows.
If we define the angles $\delta_{1}={\half}(i+j) \beta \pi$ and
$\delta_{2}={\half}(\pi-(i-j) \beta \pi)$ we can write the unitarity
equation as
\be
g^{[ij]}(\theta)g^{[ij]}(-\theta)=
{{\sin({{i \theta} \over 2}-{\pi \over 2}) \sin({{i \theta} \over 2}+
{\pi \over 2}) \sin^{2}({{i \theta} \over 2})}
\over {\sin({{i \theta} \over 2} - \delta_1) \sin({{i \theta} \over 2}+
\delta_1) \sin({{i \theta} \over 2} - \delta_2) \sin({{i \theta} \over 2}+
\delta_2)}}\ , \ \label{sufa8}
\ee
The solution for the one component case was found in \cite{shawi} where the
equation for $g(\theta)$ is:
\be
g_{\Delta}(\theta)g_{\Delta}(-\theta)={\sinh^2({\theta \over 2}) \over {\sinh^2
({\theta \over 2}) + \sin^2(\Delta \pi)}}\ , \ \label{sufa88}
\ee
where $\Delta$ is some parameter. The solution is then
\be
g_{\Delta}(\theta)={\sinh({\theta \over 2}) \over {\sinh({\theta \over 2}) +
i \sin({\Delta \pi})}} {\rm  exp}\left(i \int_{0}^{\infty}{dt \over t}
{{\sinh(\Delta t) \sinh((1-\Delta)t)} \over {\cosh^2({t \over 2})
\cosh(t)}} \sin({t \theta \over \pi})\right ) \ . \ \label{sufa9}
\ee
We can now find a solution for (\ref{sufa8}) by writing
a product of three functions of the form (\ref{sufa9}),
\be
 \ g^{[ij]}(\theta)={g_{\Delta_1}(\theta)
                     g_{\Delta_2}(\theta) \over
                     g_{\Delta_3}(\theta)}\ , \ \label{sufa10}
\ee
where $\Delta_1={\half}(i+j)\beta$, $\Delta_2={\half}(1 - (i-j)\beta)$ and
$\Delta_3={\half}$. This is the form of $g^{[ij]}(\theta)$ that we shall use
in later sections.

Before closing this section, we remark that the $S$-matrix $S^{[ij]}_{BF}$
in (\ref{sufa5}) is related to the hatted matrix $\widehat{S}^{[ij]}_{BF}$
of reference \cite{schoutens1} by $S_{BF} = \Pi \widehat{S}^{ij]}_{BF}$,
where $\Pi$ is an ordinary (not graded) permutation. Making this
choice (rather than working with $\Pi_{\rm graded}$) effectively
compensates for minus signs that would be induced by the fermionic
nature of the particles $f_i$, and it will allow us to treat all
particles as bosons when performing the TBA. Note however that this
choice breaks the manifest supersymmetry of our formalism.

% section III
\section{Some aspects of the TBA}

In the previous section we have recalled the conjecture, made
by one of us some five years ago, that the supersymmetric
and factorizable $S$-matrices (\ref{sufa3})--(\ref{sufa5})
are actually the true $S$-matrices for the perturbed conformal
field theories (\ref{sufa1}). This claim was made on the
basis of the compatibility of the set of conserved quantities
(which can be derived in the perturbed CFT framework)
and the bound-state structure that is implied by the proposed
$S$-matrix. One way to obtain further support for the
conjectured identification is by using the TBA. In brief,
the TBA makes it possible to express some of the UV CFT data
(such as the central charge) in terms of the $S$-matrix data,
providing a non-trivial check on the correctness of the proposed
$S$-matrix. Before we come to this analysis, we shall in this
section briefly review the TBA for diagonal and non-diagonal
$S$-matrices.

If we want to know the energy levels of a free particle on a circle
of radius $R$, we just have to send the particle on a round trip
around the circle and impose a matching condition for the wave function.
This will constrain the possible momenta and from that we can compute
the energy levels. In an equation this reads simply
\be
e^{i p 2 \pi R}=1 \ . \ \label{tba1}
\ee
{}From here we obtain $p_{n}=n/R$ for some integer $n$ and
energy levels given by $E_{n}=p_{n}^2/2m=n^{2}/2mR^2$.

Integrable models are not free of course, but due to the fact that multi-
particle scattering can be studied two particles at a time, and that this
scattering will be purely elastic, we obtain enormous simplifications and
the parallel with a free particle on a circle is very appropriate.
The only input we need is the exact $S$-matrix and the mass spectrum of our
theory.  Let us explain the case where the $S$-matrix is diagonal first, just
to make the concepts more clear. A more complete discussion of these matters
can be found in \cite{zama2,klamel1}.

\vspace{5mm}

{\it 3.1 The diagonal case}

The idea of the TBA is very simple physically. We consider a situation
where $M$ particles are put on a circle of lenght $L=2\pi R$. We then
take one particle, of mass $m_{i}$ and rapidity $\theta_{i}$, and send
it on a round
trip around the circle. Since we the $S$-matrix is diagonal we will
simply pick up a phase every time our particle meets one of the remaining
$M-1$ particles. The wave function should come back to its original value
after the particle has completed its trip and this implies
\be
e^{i m_{i} \sinh(\theta_{i}) L}\prod_{j\neq i} S^{[ij]}
(\theta_{ij})=1 \ . \ \label{tba2}
\ee
Taking the logarithm of this equation and going to the thermodynamic
limit (\hbox{$M \rightarrow \infty$}, \hbox{$L \rightarrow \infty$},
with $M/L$ constant) we obtain the following equation
\be
2 \pi P_{i}(\theta)=m_{i} L \cosh(\theta)+\sum_{j} \int d \theta^{'}
\rho_{j}(\theta^{'}) \phi_{i j}(\theta-\theta_{j})\ , \ \label{tba3}
\ee
where $P_{i}(\theta)$ is the density of available rapidities for particles of
type $i$, and $\rho_i(\theta)$ is number of such rapidities that are actually
occupied. The phase shift $\phi_{i j}(\theta)$ is
given by
\be
\phi_{i j}(\theta)= -i {\partial \ln S^{[ij]}(\theta) \over \partial
\theta}\ . \ \label{tba4}
\ee

Let us define the pseudo-energy levels $\epsilon_{i}(\theta)$ by
\be
{{\rho_i}(\theta) \over P_{i}(\theta)}= {e^{-\epsilon_{i}(\theta)}
\over (1+e^{-\epsilon_{i}(\theta)})}\ . \ \label{tba5}
\ee
The $\rho_i(\theta)$ are such as to minimize the free energy.
{}From this, we get an equation for the $\epsilon_{i}(\theta)$
\be
\epsilon_{i}(\theta)=m_i R \cosh(\theta) -\sum_j \int {d \theta' \over 2 \pi}
\phi_{ij}(\theta - \theta') \ln(1+e^{-\epsilon_{j}(\theta')})\ ,
\label{tba6}
\ee
The total energy of our system can be written in terms of
$\epsilon_{i}(\theta)$ according to
\be
E(R)=-\sum_i {m_i \over 2 \pi} \int d\theta \cosh(\theta) \ln(1+
e^{-\epsilon_i(\theta)})
\ , \
\label{tba7}
\ee
We shall call the equations (\ref{tba6}), (\ref{tba7}) the TBA
equations of the scattering theory.

{}From the TBA equations we can find numerically the Casimir energy for
any $R$. In the UV limit we can do even better and use the amazing
Roger's dilogarithm identities to derive closed-form expressions
(an explicit formula will be given in the next subsection).

\vspace{5mm}

{\it 3.2 The non-diagonal case}

In this case we have conceptually the same situation.
The difference is that now a particle, when taking a round trip,
scatters non-diagonally with all the others so that the
periodic boundary conditions lead to an equation involving a transfer
matrix of size $2^M\times 2^M$. This transfer matrix can
be written as
\bea
\lefteqn{
\left( T_{st}^{(j|j_1j_2\ldots j_M)}
   \right)_{s_1s_2\ldots s_M}^{t_1t_2\ldots t_M}
(\theta|\theta_1 \theta_2 \ldots \theta_M) =}
\nonu
&& \quad \sum_{\{r_l\}}
         (S^{[jj_1]})_{s s_1}^{r_1 t_1}(\theta-\theta_1)
         (S^{[jj_2]})_{r_1 s_2}^{r_2 t_2}(\theta - \theta_2) \ldots
         (S^{[jj_M]})_{r_{M-1} s_M}^{t t_M}(\theta -\theta_M)
\ , \ \label{tba11}
\eea
In this notation, the indices $j$ and $j_l$ denote the different
multiplets ($j, j_l = 1,2, \ldots, n$) and the indices $r_l$, $s_l$
and $t_l$ denote the two states $b$ and $f$ within each multiplet.
The periodic boundary condition is now expressed as
(see, for example, \cite{fenit1})
\be
e^{im_{j_l} \sinh(\theta_l) L}
\sum_s \left( T_{ss}^{(j_l|j_1j_2 \ldots j_M)} \right)
(\theta_l|\theta_1,...\theta_M) \psi = - \psi \ , \qquad l=1,2,\ldots,M \ ,
\label{tba10}
\ee
where $\psi$ is a $2^M$-component wave-function.

To make further use of this equation, we have to diagonalize
the transfer matrix, or at least find some relationship
among the eigenvalues that will allow us to write the TBA equations.
In general, the problem of diagonalizing a transfer matrix is extremely
difficult and many techniques have been developed to handle such problems
\cite{baxter}. We will, in the next section, use a method developed by
B.~Felderhof \cite{feld} to diagonalize the transfer matrix for the
eight-vertex model in the case where the Boltzmann weights satisfy the
free fermion condition. Remarkably, the resulting TBA equations can
be cast in a form identical to (\ref{tba6}), (\ref{tba7}), this time for
a set of particles that the contains the original set {\it plus}
an additional particle of mass zero (see section V). This shows
that the form (\ref{tba6}), (\ref{tba7}) of the TBA equations is
universal, applying both to diagonal and non-diagonal cases.

Returning to the Casimir energy, it can be shown that the UV limit is
given by
\be
E_0 =
E(mR \rightarrow 0) \sim
- {1 \over {\pi R}} \sum_i \left [ {\cal L}({x_i \over {1+x_i}})
- {\cal L}({y_i \over {1+y_i}}) \right ]\ , \ \label{tba8}
\ee
where ${\cal L}(x)$ is the Roger's dilogarithm function, which is defined by
\be
{\cal L}(x)= -{1 \over 2} \int^x_0 dt \left[{\ln(t) \over {1-t}} +
{\ln(1-t) \over t} \right]\ , \ \label{tba9}
\ee
and $x_i=e^{-\epsilon_i(0)}$ and $y_i=e^{-\epsilon_i(\infty)}$.
The $x_i$, $y_i$ are the solutions of
\be
x_i=\prod_j (1+x_j)^{N_{i,j}} \ , \qquad
y_{i'}=\prod_{j'} (1+y_{j'})^{N_{i',j'}} \ , \ \label{tba13}
\ee
where the $N_{i,j}$ are given by
\be
N_{i,j}={1 \over {2 \pi}} \int_{-\infty}^{\infty} d\theta \,
\phi_{ij}(\theta)\ . \ \label{tba14}
\ee
The primed indices $i'$ refer to the massless particles only.

% section IV
\section{Diagonalizing the Transfer Matrix}

As we have seen in the previous section, we need to gain control over
the eigenvalues of the trace of the transfer matrix
$T(\theta|\theta_1,...,\theta_N)$. Notice that this is
a inhomogeneous transfer matrix, depending on all the differences
$(\theta-\theta_l)$. Due to the Yang-Baxter equation, $\tr[T(\theta)]$
and $\tr[T(\theta')]$ commute for any $\theta$ and $\theta'$.
This means that the eigenvectors of $\tr[T(\theta)]$ are
$\theta$-independent.  The eigenvalues will of course depend
on $\theta$ and this dependence is precisely what we need to
know to do the TBA.

In section II, we mentioned that the non-diagonal part of the
$S$-matrix, called $S^{[ij]}_{BF}$, can be derived by
imposing supersymmetry and the Yang-Baxter equation.
A remarkable property is that the elements of
$S_{BF}^{[ij]}$, when interpreted as the Boltzmann weights of a lattice
statistical mechanics model, satisfy the {\it free fermion condition}.
Explicitly this means that, for $S^{[ij]}_{BF}$ given by
\be
S^{[ij]}_{BF} =\left(\begin{array}{cccc}
        a_{+}  &  0  &  0  &  d \\
        0      &  b_{+}  &  c  &  0 \\
        0      &  c  &  b_{-}  &  0 \\
        d      &  0  &  0  &  a_{-}
\end{array}\right)\ \ , \ \label{dtm1}
\ee
the $a_{+}$, $a_{-}$, $b_{+}$, $b_{-}$, $c$ and $d$ satisfy
\be
a_{+}a_{-}+b_{+}b_{-}=c^2+d^2 \ . \ \label{dtm2}
\ee
It can be proved that an eight-vertex model
that satisfies this condition can be written as an $XY$ model with a
magnetic field. The magnetic field is given in tems of the Boltzmann
weights
\be
H={a_{-}^2+b_{+}^2-a_{+}^2-b_-^2 \over {2(a_{+}b_{-}+a_{-}b_{+})}}\ . \
\label{dtm3}
\ee
Using the results from section II we find that $H=-1$ for all $i$, $j$.
This is the critical point of the $XY$ model.

For the determination of the eigenvalues of $\tr[T(\theta)]$ we will
follow a method given by Felderhof \cite{feld}.
We shall explain the idea of this method and apply it to our case.
The strategy will be to find an inversion relation for $\tr[T(\theta)]$
which can then be used to write a functional equation for the
eigenvalues. It will turn out that, rather miraculously, the analysis
for general $n$ can be done by extending the analysis for $n=1$,
which was presented in \cite{ahn}. Our presentation in this section
therefore follows \cite{ahn} rather closely.

It is not difficult to see that the $S$-matrix
(\ref{sufa5}) satisfies the following relation
\be
S_{BF}^{[ij]}(\theta+i \pi)=-{g^{[ij]}(\theta+i\pi) \over {g^{[ij]}(\theta)}}
\left(\begin{array}{cccc}
      a_- & 0 & 0 & c \\
      0 & -b_- & -d & 0 \\
      0 & -d & -b_+ & 0 \\
      c & 0 & 0 & a_+
      \end{array}\right)\ \ , \ \label{dtm4}
\ee
where we are using a obvious notation. If we make a similarity
transformation by means of a matrix $M$ (acting on the
indices $r=b,f$)%
 \footnote{Notice that the indices $r$ and $s$
           on the two-body $S$-matrix $(S_{BF}^{[ij]})_{rs}^{r's'}$
           are not on equal footing in the transfer matrix
           (\ref{tba11}). In the usual terminology, the
           $r$ are called matrix indices and the
           $s$ are quantum indices.}
given by
\be
M=\left(\begin{array}{cc}
        0 & 1 \\
        1 & 0
        \end{array}\right)\ \ , \ \label{dtm5}
\ee
we obtain the following $\widetilde{S}_{BF}^{[ij]}$
\be
\widetilde{S}_{BF}^{[ij]}(\theta)=-{g^{[ij]}(\theta+i\pi) \over {g^{[ij]}
(\theta)}}
\left(\begin{array}{cccc}
      -b_+ & 0 & 0 & -d \\
       0 & a_+ & c & 0 \\
       0 & c & a_- & 0 \\
      -d & 0 & 0 & -b_-
      \end{array}\right)\ \ . \ \label{dtm6}
\ee
It is easy to see that $\widetilde{S}_{BF}^{[ij]}$ again satisfies the
free fermion condition.

Returning to the transfer matrix (\ref{tba11}), let us write
$S_l = S_{BF}^{[jj_l]}(\theta-\theta_l)$, etc., so that
$\tr[T(\theta)]=\tr[S_1 S_2 ... S_M]$ and similarly
$\tr[T(\theta+i\pi)] = \tr[\widetilde{T}(\theta)] =
\tr[\widetilde{S}_1 \widetilde{S}_2 ... \widetilde{S}_M]$.
The product $\tr[T(\theta)] \, \tr[T(\theta+i\pi)]$ can then
be written as
\be
\tr[T(\theta)] \, \tr[T(\theta+i\pi)]
= \tr \left[ (S_1 \otimes \widetilde{S}_1) (S_2 \otimes \widetilde{S}_2)
  \ldots (S_N \otimes \widetilde{S}_N) \right]\ , \ \label{dtm7}
\ee
where the notation $\otimes$ denotes a matrix product of
the $s$-indices and a tensor product on the $r$-indices.
It can be proved that there is a similarity transformation with
a constant ($\theta$-independent) $4 \times 4$ matrix $X$
(acting on the indices $r_l$, $\tilde{r}_l$) that puts
$S_l \otimes \widetilde{S}_l$ in a triangular form.
We do not give the explicit form of $X$ here, but mention the
important fact that it depends on $j$, $j_l$ through
a parameter $\phi$, given by
\be
\tanh(\phi)={2 c d \over {a_+ b_+ + a_- b_-}}\ . \ \label{dtm8}
\ee
In terms of the elements of $S_{BF}^{[jj_l]}$ we get
$\tanh(\phi) = -\alpha \, m_j$, i.e, it only depends on the index
$j$, which is common to all factors in the transfer matrices.
We can thus use one and the same $X$ to put all the
$S_l \otimes \widetilde{S}_l$ in triangular form, and the total
trace $\tr[T(\theta)] \, \tr[T(\theta+i\pi)]$ is invariant
under this operation. We find
\be
- S_{BF}^{[jj_l]}(\theta-\theta_l)
  \otimes  \widetilde{S}_{BF}^{[jj_l]}(\theta-\theta_l) \sim
\left(\begin{array}{cccc}
      M_+ & * & * & * \\
      0 & \pm F_- & * & * \\
      0 & 0 & \pm F_+ & * \\
      0 & 0 & 0 & M_-
      \end{array}\right)\ \ , \ \label{dtm9}
\ee
where $M_+$, $M_-$, $F_+$ and $F_-$ are given by

\bea
M_+=&&a_+a_- -d^2\ , \quad  M_-=a_+a_--c^2\ , \
\nonu
F_{\pm}=&&\pm \left( \sinh^2(\phi) a_{\pm}b_{\pm}
                    + \cosh^2(\phi)a_{\mp}b_{\mp}
                    \mp 2 \sinh(\phi) \cosh(\phi)cd \right) ,
\ \label{dtm10}
\eea
which can be rewritten as
\bea
M_+(\theta)=&&{-4 g_2(\theta) \over {\sinh^2(\theta)}}
\cosh({{\theta + i\beta_- \pi} \over 2})
\cosh({{\theta - i\beta_- \pi} \over 2})
\sinh({{\theta + i\beta_+ \pi} \over 2})
\sinh({{\theta - i\beta_+ \pi} \over 2})\ , \
\nonu
M_-(\theta)=&&{-4 g_2(\theta) \over {\sinh^2(\theta)}}
\sinh({{\theta + i \beta_- \pi} \over 2})
\sinh({{\theta - i \beta_- \pi} \over 2})
\cosh({{\theta + i \beta_+ \pi} \over 2})
\cosh({{\theta - i \beta_+ \pi} \over 2})\ ,
\nonu
F_+(\theta)=&&{-4 g_2(\theta) \over {\sinh^2(\theta)}}
\sinh({{\theta - i \beta_- \pi} \over 2})
\cosh({{\theta + i \beta_- \pi} \over 2})
\sinh({{\theta - i \beta_+ \pi} \over 2})
\cosh({{\theta + i \beta_+ \pi} \over 2})\ ,
\nonu
F_-(\theta)=&&{-4 g_2(\theta) \over {\sinh^2(\theta)}}
\cosh({{\theta - i \beta_- \pi} \over 2})
\sinh({{\theta + i \beta_- \pi} \over 2})
\cosh({{\theta - i \beta_+ \pi} \over 2})
\sinh({{\theta + i \beta_+ \pi} \over 2})\ , \label{dtm13}
\eea
where $\beta_{\pm} = (j_l \pm j) \beta$ and
$g_2(\theta) = g^{[jj_l]}(\theta) \, g^{[jj_l]}(\theta +i \pi)$.
The entries of the matrix (\ref{dtm9}) act diagonally on the indices
$s_l$, with the $\pm$ sign equal to $+1$ ($-1$) if total in-state
is bosonic (fermionic).

Given this triangular form for $S_l \otimes \widetilde{S}_l$,
it is very easy to compute the eigenvalues of the product
$\tr[T(\theta)] \, \tr[T(\theta+i \pi)]$. Denoting
the eigenvalues of $\tr[T(\theta)]$ by $\Lambda(\theta)$,
we find
\bea
\lefteqn{
\Lambda(\theta)\Lambda(\theta+i\pi)=}
\nonu
&& \quad (-1)^M
 \left( {\prod_{l=1}^M M_+(\theta-\theta_l)}
         + {\prod_{l=1}^M M_-(\theta-\theta_l)}+
 F \, \left[{\prod_{l=1}^M F_+(\theta-\theta_l)}
          + {\prod_{l=1}^M F_-(\theta-\theta_l)} \right]
 \right) , \
\label{dtm14}
\eea
where $F$ is $+1$ for a bosonic state and $-1$ for a
fermionic eigenstate. Notice that the dependence of $M_\pm$ and
$F_\pm$ on $j$, $j_l$ has been suppressed in the notation.
In terms of reduced eigenvalues $\lambda(\theta)$, defined by
\be
 \Lambda(\theta) = \prod_{l=1}^M 2 \,
  {g^{[jj_l]}(\theta-\theta_l) \over \sinh(\theta-\theta_l)}
  \, \lambda(\theta) \ ,\label{dtm14a}
\ee
the equation (\ref{dtm14}) can be rewritten in the following
factorized  form
\bea
\lambda(\theta)\lambda(\theta+i\pi)=
\left[ F \prod_{k=1}^n \prod_{l_k=1}^{M_k}
 \cosh({\theta-\theta_{l_k}-i(k-j)\beta \pi \over 2})
 \sinh({\theta-\theta_{l_k}+i(k+j)\beta \pi \over 2})
 \right.
\nonu
\left. +\prod_{k=1}^n \prod_{l_k=1}^{M_k}
 \sinh({\theta-\theta_{l_k}-i(k-j)\beta \pi \over 2})
 \cosh({\theta-\theta_{l_k}+i(k+j)\beta \pi \over 2})
 \right]
\nonu
\times \left[ F\prod_{k=1}^n \prod_{l_k=1}^{M_k}
 \cosh({\theta-\theta_{l_k} + i(k-j)\beta \pi \over 2})
 \sinh({\theta-\theta_{l_k} - i(k+j)\beta \pi \over 2})
 \right.
\nonu
\left. \; + \prod_{k=1}^n \prod_{l_k}^{M_k}
 \sinh({\theta-\theta_{l_k}+i(k-j)\beta \pi \over 2})
 \cosh({\theta-\theta_{l_k}-i(k+j)\beta \pi \over 2})
 \right] ,
\label{dtm15}
\eea
where we assumed that out of the $M$ particles on our circle,
$M_k$ are of type $k$, $\sum_k M_k = M$. Recall that the
auxiliary particle that makes the roundtrip is of type $j$.

The $\lambda(\theta)$ are $2 \pi i$ periodic meromorphic functions,
and as such they are completely determined by their zeros and poles.
Let us denote by $z^+$ the zeros of the first factor in the
rhs of (\ref{dtm15}) and by $z^-$ those of the second factor.
They satisfy the equations
\be
\prod_{k=1}^n \prod_{l_k=1}^{M_k}
{\tanh({z^+ -\theta_{l_k}- i(k-j)\beta \pi \over 2}) \over
 \tanh({z^+ -\theta_{l_k}+ i(k+j)\beta \pi \over 2})}= -F \ , \qquad
\prod_{k=1}^n \prod_{l_k=1}^{M_k}
{\tanh({z^- -\theta_{l_k}+ i(k-j)\beta \pi \over 2}) \over
 \tanh({z^- -\theta_{l_k}- i(k+j)\beta \pi \over 2})}= -F \ .
\label{dtm16}
\ee
Denoting by $x_m$ the {\em real} solutions of the equation
\be
\prod_{k=1}^n \prod_{l_k=1}^{M_k}
{\tanh({x -\theta_{l_k}- i k \beta \pi \over 2}) \over
\tanh({x -\theta_{l_k}+ i k \beta \pi \over 2})}= -F \ , \qquad
\label{dtm17}
\ee
one may check that the following $z^\pm$ satisfy (\ref{dtm16})
\be
     z_{\epsilon^+,m}^+ = x_m - i j \beta \pi
     + i {\scriptsize{{1-\epsilon^+ \over 2}}} \pi \ ,
\qquad
     z_{\epsilon^-,m}^- = x_m + i j \beta \pi
     + i {\scriptsize{{1-\epsilon^- \over 2}}} \pi \ ,
\label{dtm17a}
\ee
Out of each pair each of zero's $z_{\epsilon^\pm,m}^\pm$,
$\epsilon^\pm=-1,1$, one will come from $\lambda(\theta)$ and the other
from $\lambda(\theta+\pi i)$, which gives us two choices.
It turns out that the choices to be made for the two factors
(i.e. the choices for $\epsilon^+$ and $\epsilon^-$)
are correlated (see, for example, \cite{ahn}), so that for
each $m$ there are two possible factors contributing to
$\lambda(\theta)$
\be
   \sinh({\theta-z_{\epsilon,m}^+ \over 2})  \,
   \sinh({\theta-z_{-\epsilon,m}^- \over 2}) \ , \quad \epsilon = -1,1 \ .
\label{dtm17b}
\ee
We thus arrive at the following expression for the eigenvalues
\be
\lambda_{\{x_m,\epsilon_m\}}(\theta) =
 {\rm Constant} \times \prod_m
 \sinh({\theta-x_m \over 2} + i \epsilon_m {j \beta \pi \over 2})
 \cosh({\theta-x_m \over 2} - i \epsilon_m {j \beta \pi \over 2}) \ ,
\label{dtm18}
\ee

The counting of the eigenvalues (there should be $2^M$ of
them), works out as follows. There are $M-1$ solutions $x_m$
in the sector $F=-1$, which, with freely chosen $\epsilon_m$,
gives the correct number of $2^{M-1}$ eigenvalues in that
sector. In the sector $F=1$ there are $M$ solutions $x_m$, and
in order to obtain the correct number of $2^{M-1}$ eigenvalues
there, we need one constraint on the allowed values of
$\epsilon_m$. Explicit results for some small values of $M$
suggest that this constraint takes the form
$\prod_m \epsilon_m = 1$.

The expressions (\ref{dtm14a}) and (\ref{dtm18}) for the
eigenvalues $\Lambda(\theta)$ will be important
ingredients in the TBA analysis, which we present in the next
section.

% section V
\section{The Thermodynamic Bethe Ansatz}

{\it 5.1 The TBA system}

In order to do the thermodynamics we have a large number $M$ of particles
on a circle of lenght $L$. We send one of the particles,
of type $j$ and rapidity $\theta$, on a round trip
and scatter it off as it moves, one particle at a time due to integrability.
When it comes back to the original point, we impose a periodic boundary
condition. From this condition we may derive the allowed
rapidity configurations, and from there the thermodynamics.

As we have seen before, the full $S$-matrix can be written as a scalar
factor $S^{[ij]}_B(\theta)$, given by (\ref{sufa4}), times the
$4 \times 4$ matrix $S_{BF}^{[ij]}(\theta)$. We will denote
the product $S_B^{[ij]}(\theta) g^{[ij]}(\theta)$ by $Z^{[ij]}(\theta)$.
The periodic boundary condition (\ref{tba10}) then takes the following
form
\be
e^{im_j \sinh(\theta) L}
{\prod_{k=1}^n \prod_{l_k=1}^{M_k}}
{Z^{[j k]}(\theta-\theta_{l_k}) \over
 {\sinh(\theta-\theta_{l_k})}}
\lambda_{\{x_m,\epsilon_m\}}(\theta)= -1 \ . \
\label{etba1}
\ee

We denote by $\rho_k(\theta)$ the density of occupied states and by
$P_k(\theta)$ the density of available states for particles of type
$k$, $k=1,2,\ldots,n$. Similarly, we introduce the symbol $P_0(\theta)$
for the density distribution of solutions $x_m$ to the equation
(\ref{dtm17}). This is further split as $P_0(\theta) = \rho_0(\theta)
+\bar{\rho}_0(\theta)$ where the densities $\rho_0$, $\bar{\rho}_0$
refer to those $x_m$ for which the corresponding $\epsilon_m = +1,-1$,
respectively. Notice that $\rho_0(\theta)$ can be viewed as a density
distribution for a new type of ``particle'' with label $0$. This
explains that the final TBA system will be in terms of $n+1$
rather than $n$ types of particles.

{}From (\ref{dtm17}) we obtain the density distribution of the
available rapidities for the ``particles'' of type $0$
\be
2 \pi P_0(\theta) =
\sum_{k=1}^n
\int d \theta' \rho_k(\theta') {\partial \over {\partial \theta}}
{\rm Im} \ln \left({\tanh({1 \over 2}(\theta-\theta'-i k \beta \pi)) \over
                   {\tanh({1 \over 2}(\theta-\theta'+i k \beta \pi))}}
             \right) \ . \ \label{etba2}
\ee
Notice that the ``mass term'', which in general occurs on the
r.h.s. of TBA equations of this sort, is absent. This means that
the auxiliary particles of type $0$ should be viewed as massless.

Taking the derivative of the imaginary part of the logarithm of
(\ref{etba1}) leads to
\bea
2 \pi P_j(\theta)
= m_j \cosh(\theta) L
  + \int d \theta' \left[ \sum_{k=1}^n
    \left( \rho_k(\theta') {\partial \over {\partial \theta}}
    {\rm Im}  \ln \left({Z^{[jk]}(\theta-\theta') \over
    {\sinh(\theta-\theta')}}\right) \right) \right.
\nonu
    \left. + \rho_0(\theta') {\partial \over {\partial \theta}}
    {\rm Im} \ln (\lambda^j_0(\theta-\theta'))
           + \bar{\rho}_0 (\theta') {\partial \over {\partial \theta}}
    {\rm Im} \ln (\bar{\lambda}^j_0(\theta-\theta')) \right]
\ , \ \label{etba3}
\eea
where $\lambda^j_0$, $\bar{\lambda}^j_0$ are defined by
\bea
&& \lambda^j_0
  = \sinh({\half}(\theta+ij\beta \pi))\cosh({\half}(\theta-ij\beta \pi))
\nonu
&& \bar{\lambda}^j_0
  = \sinh({\half}(\theta - ij\beta \pi))\cosh({\half}(\theta + ij\beta \pi))
\ ,
\label{etba3a}
\eea

Let us define some quantities that will make the notation lighter.
Define $\Phi_z^{[jk]}(\theta)$ and $\Phi^j(\theta)$ as
\bea
&& \Phi_z^{[jk]}(\theta)
 = {\partial \over {\partial \theta}}
  {\rm Im} \ln \left({Z^{[jk]}(\theta) \over {\sinh(\theta)}}\right)
\nonu
&& \Phi^j(\theta)=2 {\partial \over {\partial \theta}}
               {\rm Im} \ln (\lambda_0^j(\theta))
              = - 2 {\partial \over {\partial \theta}}
                  {\rm Im} \ln (\bar{\lambda}_0^j(\theta))
 \ . \ \label{etba4}
\eea
Notice that the same quantity $\Phi^k(\theta)$ occurs in the r.h.s.
of the equation (\ref{etba2}).

Using $P_0(\theta) = \rho_0(\theta) + \bar{\rho}_0(\theta)$ we can
now eliminate $\bar{\rho}_0(\theta)$ from (\ref{etba2}), (\ref{etba3})
to obtain
\bea
\lefteqn{2 \pi P_j(\theta) = m_j \cosh(\theta) L}
\nonu
&&  + \sum_{k=1}^n \int d \theta'
         \left( \rho_k (\theta') \left( \Phi_z^{[jk]}(\theta-\theta')
    - {1 \over 2} (\Phi^j \star \Phi^k)(\theta-\theta') \right)
    + \rho_0(\theta') \Phi^j(\theta-\theta') \right) \ , \
\label{etba5}
\eea
where we introduced the convolution
\be
(\Phi^j \star \Phi^k)(\theta)= { 1 \over 2\pi}
 \int_{-\infty}^{+\infty} d \theta' \,
 \Phi^j(\theta-\theta') \Phi^k(\theta') \ . \ \label{convo}
\label{etba5a}
\ee

We can now vary the $\rho_i(\theta)$, $\rho_0(\theta)$ so as
to minimize the free energy. In terms of the quasi-particle energies
$\epsilon_a(\theta)$, $a=0,1,2,\ldots, n$, defined as
\be
{\rho_a (\theta) \over {P_a(\theta)}}
={ e^{ - \epsilon_a (\theta)} \over 1 + e^{- \epsilon_a (\theta)} }\ , \
\label{etba5b}
\ee
this gives the equations

\bea
\epsilon_j(\theta)
&=&
  m_j L \cosh(\theta)
  - \sum_{k=1}^n
     \left( (\Phi_z^{[jk]} - {\half}
            (\Phi^{j} \star \Phi^{k})) \star \ln(1+e^{-\epsilon_{k}})
     \right) (\theta)
\nonu
&& + (\Phi^j \star \ln ( 1 + e^{-\epsilon_0} ) ) (\theta)
\nonu
\epsilon_0(\theta)
&=& -\sum_{k=1}^n (\Phi^k \star
\ln(1+e^{-\epsilon_k}))(\theta)\ , \
\label{etba6}
\eea
which have the general form (\ref{tba6}). The total energy is
given by (\ref{tba7}) with the sum running over the index
$a=0,1,2,\ldots,n$.

Now that we have the TBA equations we can explore the UV limit
and obtain analytical solutions for the ground state energy.
This will be done in the next subsection.

\vspace{5mm}

{\it 5.2 The UV limit}

The UV limit provides an important test for the conjectured
$S$-matrix since we can compute the ground state energy analytically
in this limit and compare it with the value predicted by CFT:
$E_0=- {\pi \over 6R} c_{\rm eff}$, with
$c_{\rm eff} = c-12(h_{\rm min}+{\bar h}_{\rm min})$, where
$h_{\rm min}$ (${\bar h}_{\rm min}$) is the smallest conformal dimension
in the theory. In the case of unitary theories
$h_{\rm min}= {\bar h}_{\rm min}=0$, corresponding to the
identity operator, which is always present in the operator content
of a unitary CFT. In the case of non-unitary theories
we have operators with negative dimensions and the effective
central charge picks up the term $-12(h_{\rm min}+{\bar h}_{\rm min})$.

In the CFT's that we started from in section II we have
$c=c_n=-3n(4n+3)/(2n+2)$ and $ h_{\rm \rm min} =  \bar{h}_{\rm \rm min}
= - {n \over 4}$ so that $c_{\rm eff} = 3n/(2n+2)$.

To obtain the ground state energy in the UV limit from the TBA
developed in the previous section we shall work out the general
result (\ref{tba8})-(\ref{tba14}) for these particular theories.

We denote by $A_{jk}$ and $B_k$ the following basic integrals
\be
A_{jk} = \int_{-\infty}^{\infty} {{d \theta} \over {2 \pi}} \,
         \Phi_z^{jk}(\theta)
\ , \qquad
B_j = \int_{-\infty}^{\infty} {{d \theta} \over {2 \pi}} \,
         \Phi^j(\theta)\ . \ \
\label{UV2}
\ee
Using the explicit form of $\Phi^{jk}_z(\theta)$ it is easy to compute
$A_{jk}$. We have
\bea
\Phi^{jk}_z(\theta)&=&{\rm Im} {\partial \over {\partial \theta}}
\ln \left( {Z^{[jk]}(\theta) \over {\sinh (\theta)}} \right)=
\nonumber \\
&=&{\rm Im} {\partial \over {\partial \theta}} \ln
\left( { [{\hat g}_{\Delta_1}(\theta)/\sinh(\theta)]
         [{\hat g}_{\Delta_2}(\theta)/\sinh(\theta)]
   \over [{\hat g}_{\Delta_3}(\theta)/\sinh(\theta)] }
  e^{i \int_{-\infty}^{\infty}
  {dt \over t} \sin ({t\theta \over \pi}) F(t)} S^{[jk]}_B(\theta) \right) \ ,
\label{UV3}
\eea
where $F(t)$ is a smooth function that falls exponentially to $0$ at
$t \rightarrow \pm \infty$ and ${\hat g}_{\Delta}$ is the prefactor
of formula (\ref{sufa9}).
{} From equation (\ref{sufa10}) we see that the ``$g_\Delta$" piece gives
${\half} + {\half} - {\half}$. The exponential gives zero since after the
derivative with respect to $\theta$ we get an integral of a smooth
function $F(t)$ times $\sin({t\theta \over \pi})$ and by the
Riemann-Lebesgue lemma this will be zero in the limit
$\theta \rightarrow \pm \infty$. Each factor $F_\alpha$ (with
$\alpha \neq 0$) from the bosonic  $S$-matrix (\ref{sufa4}) can be
easily seen to give $-1$, and therefore we obtain
\be
\int_{-\infty}^{\infty} {d\theta \over {2\pi}}
  {\partial \over {\partial \theta}} {\rm Im}
\ln(S_B^{[jk]}(\theta))=-2k + \delta_{j,k} \ ,
 \quad j \geq k \ . \label{UV4}
\ee
Collecting these results we see that $A_{jk}=-2j+\delta_{j,k} + {\half}$
for $j\geq k$. The computation of $B_j$ is straightforward and gives
$B_j=1$. This is all we need to write the equations (\ref{tba13})
for the pseudo-energies in the UV limit
\be
x_j = (1+x_0)^{N_{0,j}}\prod_{k=1}^n(1+x_k)^{N_{j,k}} \ , \quad
x_0 = \prod_{k=1}^n(1+x_k)^{N_{0,k}} \ , \quad y_0=1 . \ \label{UV5}
\ee
with $x_j=e^{-\epsilon_j(0)}$, $x_0=e^{-\epsilon_0(0)}$,
$y_j=e^{-\epsilon_j(\infty)}$ and $y_0=e^{-\epsilon_0(\infty)}$.
Note that due to the presence of the mass term, in the limit
$\theta \rightarrow \infty$  the $\epsilon_j(\theta)$ diverge
and this implies directly that $y_j=0$.
The $N_{a,b}$ are given by $N_{j,k}=A_{jk} - {\half} B_j B_k$
and $N_{0,j} = B_j$ so that
\be
N_{j,k}= -2k + \delta_{j,k} \ , \qquad
N_{0,j}=1 \ , \qquad
N_{0,0}=0 \ . \
\label{UV6}
\ee
for $j \ge k$. The other $N_{a,b}$ are obtained using the symmetry
$N_{a,b}=N_{b,a}$.

The (far from trivial) solutions for equation (\ref{UV5}) are
\be
x_j={\sin^2({\pi \over 2n+2}) \over \sin({(2j+3)\pi \over 4n+4})
\sin({(2j-1)\pi \over 4n+4})} \ , \qquad
x_0={\sin ({3 \pi \over 4n+4}) \over \sin({\pi \over 4n+4})} \ , \qquad
y_j = 0 \ , \qquad y_0=1 \ .
\label{UV7} \\
\ee
We can now use the Roger's dilogarithm technology and compute, using
equation (\ref{tba8}), the effective central charge $c_{\rm eff}$
\be
c_{\rm eff}={6 \over {\pi^2}} \left[\sum_{j=1}^n{\cal L}
\left({\sin^2({\pi \over 2n+2}) \over \sin^2({(2j+1)\pi \over 4n+4})} \right)+
{\cal L}\left(1-{1 \over 4 \cos^2({\pi \over 4n+4})} \right)-
{\cal L}\left( {1 \over 2} \right) \right]={3n \over 2n+2} \ ,
\ \label{UV8}
\ee
in agreement with the value in the original conformal field theory.

We found the solutions (\ref{UV7}) for $x_j$, $x_0$ and the
dilogarithm identity (\ref{UV8}) in the second paper of
\cite{fenit1}, which deals with TBA systems for $N\!=\!2$
supersymmetric theories. The relation between the $N\!=\!1$
and $N\!=\!2$ TBA systems is the topic of our next section.

% section VI
\section{Melzer's Folding}

In a very beautiful paper \cite{melzer1}, E.~Melzer studied a series of
new TBA systems with some mathematical applications in mind, namely, the
study of number theoretical identities of the Gordon-Andrews type. He
developed some TBA systems that provide such identities. Melzer's
obtained his ``first TBA system'' by a folding in half a TBA sytem
derived by P.~Fendley and K.~Intriligator \cite{fenit1}
for specific $N\!=\!2$ supersymmetric theories.
He then conjectured that this folded
TBA system should correspond to the $N\!=\!1$ supersymmetric
scattering theories of \cite{schoutens1}, which we discussed in
section II. Now that we have derived explicitly the
$N\!=\!1$ TBA systems we are in a position to check, and confirm,
Melzer's proposal.

Let us briefly summarize Melzer's discussion.
We consider a general TBA system
\be
E(R)= - \sum_a {m_a \over 2\pi} \int_{-\infty}^{+\infty}
{d\theta \over 2 \pi} \cosh(\theta) \ln(1+e^{-\epsilon_a(\theta)})
\ \label{fold2}
\ee
with
\be
\epsilon_a(\theta)=  m_a R \cosh(\theta) -\sum_b \left( \Phi_{a,b}
\star
\ln(1+e^{-\epsilon_b}) \right) (\theta)\ , \ \label{fold1}
\ee
where the kernels $\Phi_{a,b}$ are symmetric in $a,b$.
Let us now assume that
we have $2n$ types of particles, $a=1,2,\ldots,2n$ and that the TBA data
$m_a$, $\Phi_{a,b}$ possess the following symmetry
\be
m_a=m_{2n+1-a} \ , \qquad \Phi_{a,b}=\Phi_{2n+1-a,2n+1-b} \ .
\ee
In such a situation we can define a folded TBA system for half
the number of particles, $a=1,2,\ldots,n$,
by defining a folded kernel
\be
\Phi^{\rm folded}_{a,b} = \Phi_{a,b} + \Phi_{a,2n+1-b} \ ,
\quad a,b = 1,2,\ldots,n \ .
\label{fold}
\ee

The $N\!=\!2$ supersymmetric TBA systems of \cite{fenit1} contain
$2k-2$ massive particles and auxiliary massless species labeled
as $0$, $\bar{0}$. They correspond to massive scattering theories that
are obtained by starting from the $N\!=\!2$ minimal superconformal
field theories and perturbing by the most relevant supersymmetry
preserving operator. This corresponds to the Landau-Ginzburg
superpotential ${X^{2k} \over {2k}}-\lambda X$, $k=2,3,...$.
The masses are given by
\be
   m_a = { \sin({a\pi \over 2k-1}) \over \sin({\pi \over 2k-1}) } \ ,
   \qquad a=1,2,\ldots,2k-2 \ ,
\label{n=2masses}
\ee
so that $m_a=m_{2k-1-a}$, and the kernel $\Phi^{N=2}_{a,b}$ has the
symmetry property $\Phi^{N=2}_{a,b} = \Phi^{N=2}_{2k-1-a,2k-1-b}$
(we identify the labels $\bar{0}=2k-1$).
We can thus use (\ref{fold}) to define a folded version of the $N\!=\!2$
TBA system. According to Melzer's conjecture, this folded $N\!=\!2$ system
will be the same as the $N\!=\!1$ system that we derived in this paper.

Identifying $n=k-1$, we see that the mass spectra (\ref{intro1}) and
(\ref{n=2masses}) indeed agree. We also checked that
\be
\Phi^{N=1}_{a,b}(\theta)
= \Phi^{N=2}_{a,b}(\theta)
  + \Phi^{N=2}_{a,2n+1-b}(\theta) \ , \qquad a,b = 0,1,2,\ldots,n \ ,
\ee
where $\Phi^{N=2}_{a,b}(\theta)$ are the kernels obtained
in \cite{fenit1} and $\Phi^{N=1}_{a,b}(\theta)$ are the kernels
in our TBA system (\ref{etba6}). For the kernels $\Phi_{k,0}(\theta)$
this result is elementary while for the kernels $\Phi_{j,k}(\theta)$
some tedious cosine Fourier transforms are needed.

In the UV limit the folding relation can be expressed as
\be
N_{a,b}^{N=1}=N_{a,b}^{N=2}+N_{a,2n+1-b}^{N=2} \ , \ \label{fold4}
\ee
and we can use the equation (4.20) from Fendley and Intriligator's
later paper \cite{fenit1} to check that this is indeed an identity.

% section VII
\section{Conclusions}

In this paper we have studied the thermodynamic limit of a family of
perturbed non-unitary superconformal field theories with $N\!=\!1$ and
central charge $c= -3(4n+3)/(2n+2)$. We identified the two-body
$S$-matrices of those theories with the Boltzmann weights of an
eight-vertex model and we have seen that supersymmetry implies the
free fermion condition and the value $H=-1$ for the magnetic field
of the associated $XY$ model. This allowed us to diagonalize the
transfer matrix by Felderhof's method and to develop its TBA.
We found the algebraic equations that the TBA leads to in the UV
limit. Solving those equations led to the correct value
for the central charges. This analysis has thus confirmed the validity
of the $S$-matrices (\ref{sufa3})--(\ref{sufa5})
and, in particular, ruled out
possible ``CDD" correction factors for those $S$-matrices.

Another interesting result was the verification that the $N=1$
TBA systems obtained here can be obtained by folding in half
a series of $N\!=\!2$ supersymmetric TBA systems. This idea
was first proposed by Melzer in \cite{melzer1}.

There are several possibilities for further work. We have seen that
Melzer's proposal works and that the folding idea can indeed be
extended to TBA systems for non-diagonal $S$-matrices. It should
be possible to understand this folding directly at the level
$S$-matrices, possibly as some generalization of the non-critical
orbifold construction of P.~Fendley and P.~Ginsparg \cite{fengis}.
This will be subject of a future publication \cite{msII}.

Recently a lot of (deserved) interest has been given to $1+1$ field
theories with a boundary, i.e., defined on a half-line.
It is not very difficult to find a boundary $S$-matrix
or reflection matrix $R(\theta)$ for the theories that we discussed
in this paper. In particular, we can easily find the boundary
analogue of the Bose-Fermi $S$-matrix $S_{BF}(\theta)$,
see (\ref{sufa5}). In order to do that we have to solve the equation
\be
Q_B(\theta)R_{BF}(\theta)=R_{BF}(\theta)Q_B(-\theta) \ , \ \label{conc1}
\ee
where $Q_B(\theta)$ is the supersymmetry charge on the half-line. In
terms of the bulk supersymmetric charges $Q(\theta)$ and ${\widetilde Q}
(\theta)$ \cite{schoutens1} we have
$Q_B(\theta)=Q(\theta)+{\widetilde Q}(\theta)$. This gives
\be
Q_B(\theta)=e^{\theta \over 2} \sigma_x +
e^{-{\theta \over 2}} \sigma_y \ . \ \label{conc2}
\ee
Solving the condition (\ref{conc1}) we obtain
\be
R_{BF}(\theta)=Z(\theta)
\left(\begin{array}{cc}
     \cosh({\theta \over 2}+i{\pi \over 4}) & 0\\
      0 & \cosh({\theta \over 2}-i{\pi \over 4})
      \end{array} \right)\ \ , \ \label{conc3}
\ee
where $Z(\theta)$ is a normalization factor.
We have verified that this boundary matrix $R_{BF}(\theta)$,
together with the bulk $S$-matrix $S_{BF}(\theta)$, satisfies
the boundary Yang-Baxter equation. Clearly, we can extend the entire
analysis performed in \cite{schoutens1}
and in this paper to the boundary case and explore the physical
meaning of such results. This will be the subject of future work.

\section{Acknowledgements}

We would like to thank Ezer Melzer, Roland Koberle and Paul Fendley
and Omar Foda for very useful
discussions at the early stage of this work. M.M. was supported by a
fellowship from CNPq (Brazil) and K.S. was supported by the foundation FOM
(the Netherlands).

\end{document}